\documentstyle[12pt]{article}

\skewchar\fivmi='177
\skewchar\sixmi='177
\skewchar\sevmi='177
\skewchar\egtmi='177
\skewchar\ninmi='177
\skewchar\tenmi='177
\skewchar\elvmi='177
\skewchar\twlmi='177
\skewchar\frtnmi='177
\skewchar\svtnmi='177
\skewchar\twtymi='177
\def\@magscale#1{ scaled \magstep #1}
\font\twfvmi  = ammi10   \@magscale5 
\skewchar\twfvmi='177
\skewchar\fivsy='60
\skewchar\sixsy='60
\skewchar\sevsy='60
\skewchar\egtsy='60
\skewchar\ninsy='60
\skewchar\tensy='60
\skewchar\elvsy='60
\skewchar\twlsy='60
\skewchar\frtnsy='60
\skewchar\svtnsy='60
\skewchar\twtysy='60
\font\twfvsy  = amsy10   \@magscale5 
\skewchar\twfvsy='60

\catcode`@=11
\def\un#1{\relax\ifmmode\@@underline#1\else
	$\@@underline{\hbox{#1}}$\relax\fi}
\catcode`@=12

\let\du=\d			
\let\um=\H			

\def\d{\delta}

\def\G{\Gamma}

\def\L{\Lambda}

\font\sc=font005			
\font\ooo=circle10			
\font\ro=manfnt				
\def\kcl{{\hbox{\ro 6}}}		
\def\kcr{{\hbox{\ro 7}}}		
\def\ktl{{\hbox{\ro \char'134}}}	
\def\ktr{{\hbox{\ro \char'135}}}	
\def\kbl{{\hbox{\ro \char'136}}}	
\def\kbr{{\hbox{\ro \char'137}}}	

\def\bo{{\raise.15ex\hbox{\large$\Box$}}}		
\def\pr{\prod}						
\def\TH{{\raise.2ex\hbox{$\displaystyle \bigodot$}\mskip-4.7mu \llap H \;}}
\def\face{{\raise.2ex\hbox{$\displaystyle \bigodot$}\mskip-2.2mu \llap {$\ddot
	\smile$}}}					

\def\sp#1{{}^{#1}}				
\def\Bar#1{\overline{#1}}			
\def\leftrightarrowfill{$\mathsurround=0pt \mathord\leftarrow \mkern-6mu
	\cleaders\hbox{$\mkern-2mu \mathord- \mkern-2mu$}\hfill
	\mkern-6mu \mathord\rightarrow$}
\def\dvec#1{\vbox{\ialign{##\crcr
	\leftrightarrowfill\crcr\noalign{\kern-1pt\nointerlineskip}
	$\hfil\displaystyle{#1}\hfil$\crcr}}}		

\def\frac#1#2{{\textstyle{#1\over\vphantom2\smash{\raise.20ex
	\hbox{$\scriptstyle{#2}$}}}}}			
\def\sfrac#1#2{{\vphantom1\smash{\lower.5ex\hbox{\small$#1$}}\over
	\vphantom1\smash{\raise.4ex\hbox{\small$#2$}}}}	
\def\bfrac#1#2{{\vphantom1\smash{\lower.5ex\hbox{$#1$}}\over
	\vphantom1\smash{\raise.3ex\hbox{$#2$}}}}	
\def\afrac#1#2{{\vphantom1\smash{\lower.5ex\hbox{$#1$}}\over#2}}    

\newskip\humongous \humongous=0pt plus 1000pt minus 1000pt
\def\caja{\mathsurround=0pt}
\def\eqalign#1{\,\vcenter{\openup2\jot \caja
	\ialign{\strut \hfil$\displaystyle{##}$&$
	\displaystyle{{}##}$\hfil\crcr#1\crcr}}\,}
\newif\ifdtup
\def\panorama{\global\dtuptrue \openup2\jot \caja
	\everycr{\noalign{\ifdtup \global\dtupfalse
	\vskip-\lineskiplimit \vskip\normallineskiplimit
	\else \penalty\interdisplaylinepenalty \fi}}}
\def\li#1{\panorama \tabskip=\humongous				
	\halign to\displaywidth{\hfil$\displaystyle{##}$
	\tabskip=0pt&$\displaystyle{{}##}$\hfil
	\tabskip=\humongous&\llap{$##$}\tabskip=0pt
	\crcr#1\crcr}}

\def\ref#1{$\sp{#1)}$}

\parskip=\medskipamount			
\thicklines			    

\def\oldheadpic{				
	\setlength{\unitlength}{.4mm}
	\thinlines
	\par
	\begin{picture}(349,16)
	\put(325,16){\line(1,0){4}}
	\put(330,16){\line(1,0){4}}
	\put(340,16){\line(1,0){4}}
	\put(335,0){\line(1,0){4}}
	\put(340,0){\line(1,0){4}}
	\put(345,0){\line(1,0){4}}
	\put(329,0){\line(0,1){16}}
	\put(330,0){\line(0,1){16}}
	\put(339,0){\line(0,1){16}}
	\put(340,0){\line(0,1){16}}
	\put(344,0){\line(0,1){16}}
	\put(345,0){\line(0,1){16}}
	\put(329,16){\oval(8,32)[bl]}
	\put(330,16){\oval(8,32)[br]}
	\put(339,0){\oval(8,32)[tl]}
	\put(345,0){\oval(8,32)[tr]}
	\end{picture}
	\par
	\thicklines
	\vskip.2in}
\def\oldtitle#1#2#3#4{\oldheadpic\begin{center}\vglue.5in{\large\bf #1}\\[.6in]
	{#2}\\[.1in] {\it Department of Physics and Astronomy}\\
	{\it University of Maryland, College Park, MD 20742}\\[.6in]
	Physics Publication \#{#3}\\ {#4}\\[1.5in] {\bf Abstract}\\[.1in]
	\end{center} \begin{quotation}}			
\def\oldTitle#1#2#3#4#5#6#7{\oldheadpic\begin{center} \vglue .4in
	{\large\bf #1}\\[.4in]
	{#2}\\[.1in] {\it Department of Physics and Astronomy}\\
	{\it University of Maryland, College Park, MD 20742}\\[.1in]
	{#3}\\[.1in] {\it {#4}}\\ {\it {#5}}\\[.4in]
	Physics Publication \#{#6}\\ {#7}\\[.5in] {\bf Abstract}\\[.1in]
	\end{center} \begin{quotation}}			
\def\border{						
	\setlength{\unitlength}{1mm}
	\newcount\xco
	\newcount\yco
	\xco=-24
	\yco=12
	\begin{picture}(140,0)
	\put(\xco,\yco){$\ktl$}
	\advance\yco by-1
	{\loop
	\put(\xco,\yco){$\kcl$}
	\advance\yco by-2
	\ifnum\yco>-240
	\repeat
	\put(\xco,\yco){$\kbl$}}
	\xco=158
	\yco=12
	\put(\xco,\yco){$\ktr$}
	\advance\yco by-1
	{\loop
	\put(\xco,\yco){$\kcr$}
	\advance\yco by-2
	\ifnum\yco>-240
	\repeat
	\put(\xco,\yco){$\kbr$}}
        \put(-20,11){\tiny University of Maryland Elementary Particle
Physics University of Maryland Elementary Particle Physics University of
Maryland Elementary Particle Physics}
	\put(-20,-241.5){\tiny University of Maryland Elementary
Particle Physics University of Maryland Elementary Particle Physics
University of Maryland Elementary Particle Physics}
	\end{picture}
	\par\vskip-8mm}
\def\bordero{						
	\setlength{\unitlength}{1mm}
	\newcount\xco
	\newcount\yco
	\xco=-24
	\yco=12
	\begin{picture}(140,0)
	\put(\xco,\yco){$\ktl$}
	\advance\yco by-1
	{\loop
	\put(\xco,\yco){$\kcl$}
	\advance\yco by-2
	\ifnum\yco>-240
	\repeat
	\put(\xco,\yco){$\kbl$}}
	\xco=158
	\yco=12
	\put(\xco,\yco){$\ktr$}
	\advance\yco by-1
	{\loop
	\put(\xco,\yco){$\kcr$}
	\advance\yco by-2
	\ifnum\yco>-240
	\repeat
	\put(\xco,\yco){$\kbr$}}
	\put(-20,12){\ooo
bacdefghidfghghdhededbihdgdfdfhhdheidhdhebaaahjhhdahbahgdedgehgfdiehhgdigicba}
	\put(-20,-241.5){\ooo
ababaighefdbfghgeahgdfgafagihdidihiidhiagfedhadbfdecdcdfagdcbhaddhbgfchbgfdacfediacbabab}
	\end{picture}
	\par\vskip-8mm}
\def\headpic{						
	\indent
	\setlength{\unitlength}{.4mm}
	\thinlines
	\par
	\begin{picture}(29,16)
	\put(165,16){\line(1,0){4}}
	\put(170,16){\line(1,0){4}}
	\put(180,16){\line(1,0){4}}
	\put(175,0){\line(1,0){4}}
	\put(180,0){\line(1,0){4}}
	\put(185,0){\line(1,0){4}}
	\put(169,0){\line(0,1){16}}
	\put(170,0){\line(0,1){16}}
	\put(179,0){\line(0,1){16}}
	\put(180,0){\line(0,1){16}}
	\put(184,0){\line(0,1){16}}
	\put(185,0){\line(0,1){16}}
	\put(169,16){\oval(8,32)[bl]}
	\put(170,16){\oval(8,32)[br]}
	\put(179,0){\oval(8,32)[tl]}
	\put(185,0){\oval(8,32)[tr]}
	\end{picture}
	\par\vskip-6.5mm
	\thicklines}
\def\title#1#2#3#4{\border\headpic {\hbox to\hsize{#4 \hfill UMDEPP #3}}\par
	\begin{center} \vglue .5in {\large\bf #1}\\[.6in]
	{#2}\\[.1in] {\it Department of Physics and Astronomy}\\
	{\it University of Maryland, College Park, MD 20742}\\[1.5in]
	{\bf Abstract}\\[.1in] \end{center} \begin{quotation}}	
\def\Title#1#2#3#4#5#6#7{\border\headpic
	{\hbox to\hsize{#7 \hfill UMDEPP #6}}\par
	\begin{center} \vglue .4in {\large\bf #1}\\[.4in]
	{#2}\\[.1in] {\it Department of Physics and Astronomy}\\
	{\it University of Maryland, College Park, MD 20742}\\[.1in]
	{#3}\\[.1in] {\it {#4}}\\ {\it {#5}}\\[.5in] {\bf Abstract}\\[.1in]
	\end{center} \begin{quotation}}			
\def\endtitle{\end{quotation}\newpage}			

\def\sect#1{\bigskip\medskip \goodbreak \noindent{\bf {#1}} \nobreak \medskip}
\def\refs{\sect{References} \footnotesize \frenchspacing \parskip=0pt}
\def\Item{\par\hang\textindent}

\def\doit#1#2{\ifcase#1\or#2\fi}
\def\[{\lfloor{\hskip 0.35pt}\!\!\!\lceil}
\def\]{\rfloor{\hskip 0.35pt}\!\!\!\rceil}

\def\du#1#2{_{#1}{}^{#2}}

\def\hati{{\hat{I}}}

\def\pl#1#2#3{Phys.~Lett.~{\bf {#1}B} (19{#2}) #3}
\def\np#1#2#3{Nucl.~Phys.~{\bf B{#1}} (19{#2}) #3}

\def\pr#1#2#3{Phys.~Rev.~{\bf D{#1}} (19{#2}) #3}

\def\cmp#1#2#3{Comm.~Math.~Phys.~{\bf {#1}} (19{#2}) #3}
\def\jmp#1#2#3{Jour.~Math.~Phys.~{\bf {#1}} (19{#2}) #3}

\def\ijmp#1#2#3{Int.~Jour.~Mod.~Phys.~{\bf {#1}} (19{#2}) #3}

\def\ibid#1#2#3{{\it ibid.}~{\bf {#1}} (19{#2}) #3}

\def\ula{{\un a}}
\def\ulb{{\un b}}
\def\ulc{{\un c}}
\def\uld{{\un d}}

\def\fracmm#1#2{{{#1}\over{#2}}}
\def\gg{{\hbox{\sc g}}}

\def\frac#1#2{{\textstyle{#1\over\vphantom2\smash{\raise -.20ex
	\hbox{$\scriptstyle{#2}$}}}}}			

\def\Dot#1{\buildrel{_{_{\hskip 0.01in}\bullet}}\over{#1}}

\def\uln{{\underline n}}

\def\scst{\scriptstyle}
\def\itrema{$\ddot{\scriptstyle 1}$}
\def\Bo{\bo{\hskip 0.03in}}

\def\derx{\partial_x} \def\dery{\partial_y} \def\dert{\partial_t}
\def\Vec#1{{\overrightarrow{#1}}}
\def\.{.$\,$}

\def\grg#1#2#3{Gen.~Rel.~Grav.~{\bf{#1}} (19{#2}) {#3} }
\def\pla#1#2#3{Phys.~Lett.~{\bf A{#1}} (19{#2}) {#3}}

\def\ula{{\underline a}} \def\ulb{{\underline b}} \def\ulc{{\underline c}}
\def\uld{{\underline d}} \def\ule{{\underline e}} \def\ulf{{\underline f}}
\def\ulg{{\underline g}} \def\ulm{{\underline m}}
\def\uln#1{\underline{#1}}
\def\ulp{{\underline p}} \def\ulq{{\underline q}} \def\ulr{{\underline r}}

\def\hatm{\hat m}\def\hatn{\hat n}\def\hatr{\hat r}\def\hats{\hat s}
\def\hatt{\hat t}

\def\plpl{{+\!\!\!\!\!{\hskip 0.009in}{\raise -1.0pt\hbox{$_+$}}
{\hskip 0.0008in}}}

\def\mimi{{-\!\!\!\!\!{\hskip 0.009in}{\raise -1.0pt\hbox{$_-$}}
{\hskip 0.0008in}}}

\def\ul{\underline}
\def\un{\underline}
\def\-{{\hskip 1.5pt}\hbox{-}}

\def\kd#1#2{\d\du{#1}{#2}}
\def\fracmm#1#2{{{#1}\over{#2}}}
\def\footnotew#1{\footnote{\hsize=6.5in {#1}}}

\def\low#1{{\raise -3pt\hbox{${\hskip 1.0pt}\!_{#1}$}}}

\def\ip{{=\!\!\! \mid}}
\def\unb{{\underline {\bar n}}}
\def\upb{{\underline {\bar p}}}
\def\um{{\underline m}}
\def\up{{\underline p}}
\def\Phib{{\Bar \Phi}}
\def\Phit{{\tilde \Phi}}
\def\Phibt{{\tilde {\Bar \Phi}}}
\def\Db{{\Bar D}_{+}}
\def\gg{{\hbox{\sc g}}}
\def\nt{$~N=2$~}

\def\Dot#1{\buildrel{_{_{\hskip 0.01in}\bullet}}\over{#1}}

\def\gg{{\hbox{\sc g}}}
\def\nt{$~N=2$~}
\def\gg{{\hbox{\sc g}}}
\def\nt{$~N=2$~}
\def\tr{{\rm tr}}
\def\Tr{{\rm Tr}}
\def\mpl#1#2#3{Mod.~Phys.~Lett.~{\bf A{#1}} (19{#2}) #3}
\def\hati{{\hat i}} \def\hatj{{\hat j}} \def\hatk{{\hat k}}
\def\hatl{{\hat l}}

\begin{document}
\font\tenmib=cmmib10
\font\sevenmib=cmmib10 at 7pt 
\font\fivemib=cmmib10 at 5pt  
\font\tenbsy=cmbsy10
\font\sevenbsy=cmbsy10 at 7pt 
\font\fivebsy=cmbsy10 at 5pt  
\def\BMfont{\textfont0\tenbf \scriptfont0\sevenbf
                              \scriptscriptfont0\fivebf
            \textfont1\tenmib \scriptfont1\sevenmib
                               \scriptscriptfont1\fivemib
            \textfont2\tenbsy \scriptfont2\sevenbsy
                               \scriptscriptfont2\fivebsy}
\def\rlx{\relax\leavevmode}
\def\BM#1{\rlx\ifmmode\mathchoice
                      {\hbox{$\BMfont#1$}}
                      {\hbox{$\BMfont#1$}}
                      {\hbox{$\scriptstyle\BMfont#1$}}
                      {\hbox{$\scriptscriptstyle\BMfont#1$}}
                 \else{$\BMfont#1$}\fi}

\font\tenmib=cmmib10
\font\sevenmib=cmmib10 at 7pt 
\font\fivemib=cmmib10 at 5pt  
\font\tenbsy=cmbsy10
\font\sevenbsy=cmbsy10 at 7pt 
\font\fivebsy=cmbsy10 at 5pt  
\def\BMfont{\textfont0\tenbf \scriptfont0\sevenbf
                              \scriptscriptfont0\fivebf
            \textfont1\tenmib \scriptfont1\sevenmib
                               \scriptscriptfont1\fivemib
            \textfont2\tenbsy \scriptfont2\sevenbsy
                               \scriptscriptfont2\fivebsy}
\def\BM#1{\rlx\ifmmode\mathchoice
                      {\hbox{$\BMfont#1$}}
                      {\hbox{$\BMfont#1$}}
                      {\hbox{$\scriptstyle\BMfont#1$}}
                      {\hbox{$\scriptscriptstyle\BMfont#1$}}
                 \else{$\BMfont#1$}\fi}

\def\inbar{\vrule height1.5ex width.4pt depth0pt}
\def\sinbar{\vrule height1ex width.35pt depth0pt}
\def\ssinbar{\vrule height.7ex width.3pt depth0pt}
\font\cmss=cmss10
\font\cmsss=cmss10 at 7pt
\def\ZZ{\rlx\leavevmode
             \ifmmode\mathchoice
                    {\hbox{\cmss Z\kern-.4em Z}}
                    {\hbox{\cmss Z\kern-.4em Z}}
                    {\lower.9pt\hbox{\cmsss Z\kern-.36em Z}}
                    {\lower1.2pt\hbox{\cmsss Z\kern-.36em Z}}
               \else{\cmss Z\kern-.4em Z}\fi}
\def\Ik{\rlx{\rm I\kern-.18em k}}  
\def\IC{\rlx\leavevmode
             \ifmmode\mathchoice
                    {\hbox{\kern.33em\inbar\kern-.3em{\rm C}}}
                    {\hbox{\kern.33em\inbar\kern-.3em{\rm C}}}
                    {\hbox{\kern.28em\sinbar\kern-.25em{\rm C}}}
                    {\hbox{\kern.25em\ssinbar\kern-.22em{\rm C}}}
             \else{\hbox{\kern.3em\inbar\kern-.3em{\rm C}}}\fi}
\def\IP{\rlx{\rm I\kern-.18em P}}
\def\IR{\rlx{\rm I\kern-.18em R}}
\def\IN{\rlx{\rm I\kern-.20em N}}
\def\Ione{\rlx{\rm 1\kern-2.7pt l}}

\def\scst{\scriptstyle}
\def\itrema{$\ddot{\scriptstyle 1}$}
\def\Bo{\bo{\hskip 0.03in}}
\def\derx{\partial_x} \def\dery{\partial_y} \def\dert{\partial_t}
\def\Vec#1{{\overrightarrow{#1}}}
\def\.{.$\,$}
\def\res{\hbox{Res}\,}
\def\[{\lfloor{\hskip 0.35pt}\!\!\!\lceil~}
\def\]{~\rfloor{\hskip 0.35pt}\!\!\!\rceil}

\font\fieldfont=msbm10 at 12pt
\def\complexnumber{\hbox{\fieldfont C}}
\def\integernumber{\hbox{\fieldfont Z}}
\def\realnumber{\hbox{\fieldfont R}}
\def\naturalnumber{\hbox{\fieldfont N}}

\def\realnumber{\IR}
\def\integernumber{\ZZ}
\def\naturalnumber{\IN}
\def\complexnumber{\IC}

\def\llr{ < \!\! < \, }
\def\ggr{ > \!\! > \, }

\def\grg#1#2#3{Gen.~Rel.~Grav.~{\bf{#1}} (19{#2}) {#3} }
\def\pla#1#2#3{Phys.~Lett.~{\bf A{#1}} (19{#2}) {#3}}

\def\ula{{\underline a}} \def\ulb{{\underline b}} \def\ulc{{\underline c}}
\def\uld{{\underline d}} \def\ule{{\underline e}} \def\ulf{{\underline f}}
\def\ulg{{\underline g}} \def\ulm{{\underline m}}
\def\uln#1{\underline{#1}}
\def\ulp{{\underline p}} \def\ulq{{\underline q}} \def\ulr{{\underline r}}

\def\hatm{\hat m}\def\hatn{\hat n}\def\hatr{\hat r}\def\hats{\hat s}
\def\hatt{\hat t}

\def\plpl{{+\!\!\!\!\!{\hskip 0.009in}{\raise -1.0pt\hbox{$_+$}}
{\hskip 0.0008in}}}
\def\mimi{{-\!\!\!\!\!{\hskip 0.009in}{\raise -1.0pt\hbox{$_-$}}
{\hskip 0.0008in}}}

\def\items#1{\\ \item{[#1]}}
\def\ul{\underline}
\def\un{\underline}
\def\-{{\hskip 1.5pt}\hbox{-}}

\def\kd#1#2{\d\du{#1}{#2}}
\def\fracmm#1#2{{{#1}\over{#2}}}
\def\footnotew#1{\footnote{\hsize=6.5in {#1}}}

\def\low#1{{\raise -3pt\hbox{${\hskip 1.0pt}\!_{#1}$}}}

\def\ip{{=\!\!\! \mid}}
\def\unb{{\underline {\bar n}}}
\def\upb{{\underline {\bar p}}}
\def\um{{\underline m}}
\def\up{{\underline p}}
\def\Phib{{\Bar \Phi}}
\def\Phit{{\tilde \Phi}}
\def\Phibt{{\tilde {\Bar \Phi}}}
\def\Db{{\Bar D}_{+}}
\def\gg{{\hbox{\sc g}}}
\def\nt{$~N=2$~}

{\hbox to\hsize{October 1993\hfill UMDEPP 94--47}}\par

\begin{center}
\vglue .25in

{\large\bf Super--Lax Operator Embedded in}
\vskip 0.01in
{\large\bf Self--Dual Supersymmetric Yang--Mills Theory}$\,$\footnote{This
work is supported in part by NSF grant \# PHY-91-19746.} \\[.1in]

\baselineskip 10pt

\vskip 0.25in

Hitoshi ~NISHINO\footnote{E-Mail Address: Nishino@umdhep.umd.edu} \\[.2in]
{\it Department of Physics} \\ [.015in]
{\it University of Maryland at College Park}\\ [.015in]
{\it College Park, MD 20742-4111, USA} \\[.1in]
and\\[.1in]
{\it Department of Physics and Astronomy} \\[.015in]
{\it Howard University} \\[.015in]
{\it Washington, D.C. 20059, USA} \\[.18in]

\vskip 1.0in

{\bf Abstract}\\[.1in]
\end{center}

\begin{quotation}

{}~~~We show that the super-Lax operator for $~N=1$~ supersymmetric
Kadomtsev-Petviashvili equation of Manin and Radul in three-dimensions can
be embedded into recently developed self-dual supersymmetric Yang-Mills theory
in $~2+2\-$dimensions, based on general features of its underlying
super-Lax equation.  The differential geometrical relationship in
superspace between the embedding principle of the super-Lax operator and
its associated super-Sato equation is clarified.  This result provides
a good guiding principle for the embedding of other integrable sub-systems in
the super-Lax equation into the four-dimensional self-dual supersymmetric
Yang-Mills theory, which is the consistent background for $~N=2$~ superstring
theory, and potentially generates other unknown supersymmetric integrable
models
in lower-dimensions.

\endtitle

\def\Dot#1{\buildrel{_{_{\hskip 0.01in}\bullet}}\over{#1}}
\def\tr{{\rm tr}\,}
\def\Tr{{\rm Tr}\,}
\def\mpl#1#2#3{Mod.~Phys.~Lett.~{\bf A{#1}} (19{#2}) #3}
\def\hati{{\hat i}} \def\hatj{{\hat j}} \def\hatk{{\hat k}}
\def\hatl{{\hat l}}

\oddsidemargin=0.03in
\evensidemargin=0.01in
\hsize=6.5in
\textwidth=6.5in

\noindent {\bf 1.~~Introduction~~}
\vskip 0.1in

The conjecture [1] that self-dual
Yang-Mills (SDYM) theory in four-dimensions \hbox{$~(D=4)$}
\newline will generate (possibly all) integrable
systems in lower-dimensions has attracted much attention, not only
from purely mathematical interest, but also from the practical
applications in many useful models and systems in physics.

	It is not merely a coincidence that the recent development of
superstrings has also provided the important insight to the SDYM
theory, which is nothing but the consistent background for the $~N=2$~
superstring [2].  In the recent formulation of the open $~N=2$~ superstring
[3], it has been shown that if the supermultiplet for the background is to be
described by an irreducible supermultiplet, it must necessarily be the
$~N=4$~ self-dual {\it supersymmetric} Yang-Mills (SDSYM) theory.  The usage
of the superstring theory has provoked a new insight that the quantum
aspects of the SDSYM theory should be studied in the context of the $~N=2$~
superstring.

	In a recent series of papers [4-8], we have studied various
aspects of the SDSYM theories and self-dual supergravity theories,
and we have also shown [5-8] some explicit examples of embedding typical
supersymmetric
integrable systems into the $~D=4$~ SDSYM, such as the supersymmetric KdV
(SKdV) equations [9], supersymmetric Toda theory in $~D=2$~ or
supersymmetric Kadomtsev-Petviashvili (SKP) equation of Manin and Radul
[10] in $~D=3$, some topological field theory and
$~W_\infty\-$gravity [7].  We also showed [8] that such embeddings are also
possible for the case of Wess-Zumino-Novikov-Witten models, which are
closely related to supersymmetric integrable models or even more closely to
$~N=1$~ superstrings.  The success of these embeddings suggests that there must
be some universal principle underlying them.  Even though some ans{\" a}tze
are presented in refs.~[5,6] for explicit lower flows,
more general embedding principles for the entire integrable hierarchy seem
rather obscure.  Motivated by this observation, we try in this Letter to show
how the super-Lax operator for the $~N=1$~ SKP equation of Manin and
Radul [10], which possesses
one of the most common fundamental aspects for supersymmetric integrable
systems, can be embedded
into the SDSYM theory in $~D=4$.  Since the $~D=2$~ supersymmetric
integrable models such as SKdV equations [9] are
generated by further dimensional reduction of the $~D=3$~ SKP equation,
our result automatically applies also to the embedding of the formers.

	We first show how the super-Lax operator for
$~N=1$~ SKP equation of Manin and Radul [10] can be
embedded into the $~D=4$~ SDSYM.  For this purpose we use a generalized ring
for super-microdifferential operators\footnotew{The term
``super-pseudodifferential operator'' is also used.  In this Letter, we try
to accord with mathematical terminology in ref.~[11], as much as
possible.}~~instead of the usual finite-dimensional gauge Lie algebra.  We
next show the geometrical
significance of our embedding, relating it to what is called super-Sato
equation for a wave operator [11].  We also show the meaning
of infinite number of conserved charges in terms of our superspace
formulation.

\bigskip\bigskip
\bigskip\bigskip

\noindent {\bf 2.~~Embedding of $~N=1$~ SKP Equation~~}
\vskip 0.1in

We start with the
embedding of $~D=3,\,N=1$~ SKP equation by Manin and Radul [10] into the
$~D=4,\,N=1$~ SDSYM.  The super-Lax equation for even time flows for
the $~N=1$~ SKP hierarchy [10,11] is given
in terms of the super-microdifferential operator ~$L$:
$$ \fracmm{\partial L} {\partial t_{2n}}
= \[ ~(L^{2n})_+, ~L ~\]~~, ~~~~(t_2 \equiv x~,~~t_4 \equiv y~,
{}~~t_6\equiv t)~~,
\eqno(2.1) $$
where
$$\eqalign{& L \equiv \sum_{m=0} ^\infty u_m D^{1-m} ~~,
{}~~~u_m \equiv u_m(x,y,t,\theta) ~~,
{}~~~u_0 = 1~, ~~(D u_1) + 2u_2 = 0 ~~, \cr
&D^{-1} \equiv D\partial_x^{-1}~~,
{}~~~D\equiv \fracmm \partial{\partial\theta} + \theta
\fracmm \partial{\partial x} ~~, ~~~ D^2 \equiv
\fracmm\partial{\partial x} \equiv \partial_x ~~,  ~~~~~ \cr
&\partial_x^{-m} f(x,y,t)\equiv \sum_{n=0} ^\infty (-1)^n
\left( { {\scst m+n-1} \atop {\scst n} } \right)
{}~\left(\fracmm{\partial^n f}
{\partial x^n} \right) ~\partial_x^{-m-n}~~. \cr }
\eqno(2.2) $$
As usual [9,10], the subscript $~_+$~ denotes the projection onto the class of
terms with non-negative powers of $~D$.
The Lax operator $~L$~ is fermionic (anti-commuting).
Unless braced by parentheses, $~\partial_x^m,~D^m,
{}~\partial_x^{-m}$~ and ~$D^{-m}$~ are to be regarded as operators, like
$~\partial_x f \equiv (\partial f / \partial x) + f \partial_x$.
The symbol $~\left( {m \atop n} \right)$~ denotes the usual
binomial coefficients.
In this Letter we use first only $~t_2,\,t_4,\,t_6$~\footnotew{In general we
can choose any arbitrary three bosonic coordinates out of $~t_2,\, t_4,
\, t_6, \, \cdots$~ for embedding of
other sub-systems contained in the SKP super-Lax equation into $~D=3$.  See
eq.~(2.13).}
out of infinitely many
bosonic time variables $~t_2,\,t_4,\, \cdots$, and only $~\theta$~ out
of infinitely many fermionic variables $~\theta,\,t_1,\,t_3,\,
\cdots$~ in the formalism of ref.~[11], in order to see in particular
the $~N=1$~ SKP equation of Manin and Radul in $~D=3$~ [10], as an example.

	Since the super-Lax operator contains infinitely many fields
$~u_m$, as well as non-local operators like $~\partial_m^{-1}$,
the space of the algebra is expected to be
infinite-dimensional, and we have to generalize the concept of the usual
finite-dimensional Yang-Mills gauge group.  Intuitively we can
supersymmetrize the non-supersymmetric embedding by Schiff
[12], and consider a ``graded'' ring $~{\cal E}$~ of the
super-microdifferential operators.  This $~{\cal E}$~ is a kind of
gauge algebra, but it can be also identified with its universal
enveloping algebra with the usual
multiplication properties with gradations, and the unique splitting property
$~P = P_+ + P_-$~ into the polynomials of respectively non-negative and
negative powers of $~D$~ for an arbitrary element $~P\in{\cal E}$.
This is an intuitive description of the graded ring $~{\cal E}$~ of the
super-microdifferential operators.

	Before performing our embedding, let us specify the
ring $~{\cal E}$~ in mathematical terms more precisely.  For this purpose,
we follow the terminology in ref.~[11].  Consider the formal completion
$~{\cal S}$:
$${\cal S} \simeq {\cal A} \otimes {\realnumber} [[x,\,y,\,t,\, \theta]] ~~
\eqno(2.3) $$
which is the algebra of analytic functions on a super affine space
$~B_{\cal A}^{(1|1)}$~ of dimension $~(1|1)$~ over
$~{\cal A}$~ which is Grassmann algebra
$~\L ({\realnumber})\otimes {\IR}^2~$ generated by a three-dimensional vector
space $~{\realnumber}\otimes\IR^2$ with the respective coordinates $~x$~
and $~(y,\,t)$.  The $~\ZZ_2\-$gradation in $~{\cal A}$~ is applied only
to the first $~\IR$~ out of $~\IR \otimes \IR^2$~ with the variables
$~(x,\theta)$, where $~\theta$~ is anti-commuting (fermionic).
In terminology for physicists, this vector space is
nothing but what is called $~N=1$~ superspace in $~D=3$.\footnotew{Even
though we write $~D=3$, our base space is more
precisely $~(D=1,\,N=1~\hbox{superspace}) \otimes \realnumber ^2$,
because we have {\it no} fermionic $~\theta$-coordinates for the $~y$~ and
$~t$-coordinates.  In this sense our $~{\cal E}_{\cal S}^{(1|1)}$~ differs
from $~{\cal E}_{\cal S}^{1|1}$~ by Ueno and Yamada [11].  Relevantly,
our coordinates $~x,~y,~t$~ are all real.}~~
The $~\IR [[x,\,y,\,t,\,\theta]]$~ is the algebra
of formal power series with $~x,\,y,\,t$~ and $~\theta$.  The algebra
$~{\cal S}$~ is a super-commutative algebra
with the $~{\integernumber}_2\-$gradation:
$~{\cal S} = {\cal S}_0 \oplus {\cal S}_1$, where the suffix $~{\scst
0}$~ (or $~{\scst 1}$) denotes the bosonic (or fermionic) grading.
Let $~{\cal E}_{\cal S}^{(1|1)}$~ denote the ring of
super-microdifferential operators over $~{\cal S}$:
$$\eqalign{ {\cal E}_{\cal S}^{(1|1)} &\equiv \left\{ \sum_{-\infty < n
\llr \infty} u_n(x,\, y,\, t,\,\theta) D^n ~\Bigg|~
u_n(x,\, y,\, t,\, \theta) \in {\cal S} \right\} \cr
& = \bigcup_{m=-\infty}^\infty ~{\cal E}_{\cal S} ^{(1|1)} (m)
{}~~. \cr}
\eqno(2.4) $$
The last line is for a filtration compatible with the
$~{\integernumber}_2\-$gradation: $~{\cal E}_{\cal S}^{(1|1)} (m) = {\cal S}
[[D^{-1} ]] D^m$, whose typical element is
$$ P = \sum_{n = -\infty}^m p_n(x,\,y,\,t,\,\theta) D^n
\in {\cal E}_{\cal S}^{(1|1)}(m)~~,
{}~~~~(p_m \neq 0) ~~.
\eqno(2.5) $$
Notice also that $~{\cal E}_{\cal S}^{(1|1)} = {\cal S}(D) \oplus{\cal E}
_{\cal S} ^{(1|1)} (-1)$, where $~{\cal S} (D)$~ is the ring of
super-differential operators.
Accordingly, any element $~P \in {\cal E}_{\cal S}^{(1|1)} (m) $~ is
uniquely projected into
$$P = P_+ + P_-~~, ~~~~ P_+ \in {\cal S} (D) ~~,
{}~~~~P_-\in {\cal E}_{\cal S}^{(1|1)} (-1) ~~.
\eqno(2.6) $$
Reviewing $~L$~ in (2.2), we see that
$$L \in {\cal E}_{\cal S}^{(1|1)} (1) ~\bigcap ~({\cal E} _{\cal S}
^{(1|1)} ) _ 1 ~~,
\eqno(2.7) $$
namely the fermionic operator $~L$~ has the powers of $~D$~ between
$~-\infty$~ and ~$+1$.  Now the practical significance of these
mathematical terms is self-explanatory.

	We now come to our formulation of supersymmetric
Yang-Mills theory in $~N=1$~ superspace $~(Z^M)=(x,\,y,\,t,\,\theta)$~
with a potential superfields $~A_M(x,\,y,\,t,\,\theta)$~ regarded as a
connection superfields on the base superspace $~(Z^M)$~
with their values in the ring $~{\cal E}_{\cal S}^{(1|1)}$.
This is nothing else than a generalization of the finite-dimensional Lie
algebra used in supersymmetric Yang-Mills theory.  It is now clear that
the usual Lie ring of a
gauge Lie group is replaced by the ring $~{\cal E}_{\cal S}^{(1|1)}$~ of
super-microdifferential operators.
By construction, the Grassmann parities (gradations) for the coordinates
$~(x,\,y,\,t,\,\theta)$~ are identified with those for
$~{\cal E}_{\cal S}^{(1|1)}$.

	We now turn to our original purpose, namely the embedding
of the super-Lax operator for the SKP equation of Manin and Radul into the
SDSYM theory.  In our recent papers
[5,6], it has been shown that the system of vanishing Yang-Mills superfield
strength in $~D\le 3$~ can be obtained as a ``descendant''
theory of the $~D=4$~ SDSYM theory after appropriate dimensional reductions.
It is to be understood also that our embedding is equivalent to
a further embedding into the
$~N=2$~ superstring theory [2,3] as a more fundamental theory, since
the $~D=4$~ SDSYM fields are nothing but the consistent backgrounds for
the $~N=2$~ superstring theory [2,3].

	As a matter of fact, in the superspace formulation of the $~{\cal
E}_{\cal S}^{(1|1)}\-$ring algebra-valued supersymmetric Yang-Mills theory
above, we can set up such a system of vanishing superfield strength in
superspace.  Namely we can show that
$$ F_{A B} \equiv \[ \nabla_A,~\nabla_B \} - T\du{A B} C \nabla_C = 0 ~~,
\eqno(2.8)$$
for all the supercoordinate indices of $~{\scst A,~B,~\cdots ~=~
x,~y,~t,~\theta}$.

	Our embedding satisfying (2.8) can be performed by the ans{\"a}tze:
$$\li{& \nabla _\theta \equiv D + A_\theta \equiv L~~,
\cr
& \nabla_x \equiv \partial_x + A_x \equiv \nabla_\theta^2
\equiv L^2 ~~,
&(2.9a) \cr
&\nabla_y \equiv \partial_y + A_y \equiv \partial_y - (L^4)_+ + L^4~~, \cr
&\nabla_t \equiv \partial_t + A_t \equiv \partial_t - (L^6)_+ + L^6 ~~,
&(2.9b) \cr } $$
where $~\partial_y \equiv \partial / \partial y$~ and
$~\partial_t \equiv \partial / \partial t$.

	It is now easy to show that (2.8) holds for $~{\scst A\, B ~=~
\theta\theta,~  x\,\theta, ~ y\,\theta, ~t\,\theta, ~a\,b}$, where $~{\scst
a,~b,~\cdots~=~ x,~y,~t}$.  For example,
$$\eqalign{ F_{\theta\theta} & = \{ L,\, L\} - 2 \nabla_x \equiv 0~~, \cr
F_{x \theta} & = \[ \nabla_x ,\, \nabla_\theta \]
= \[\partial_x - (L^2) _+ + L^2 ,~L\] \cr
& = \fracmm{\partial L}{\partial x} - \[ (L^2)_+ , \, L\] = 0 ~~, \cr
F_{y\theta} & = \[ \partial_y - (L^4)_+ + L^4, \, L \] = 0 ~~, \cr
F_{x y} & = \[ L^2 ,\, \partial_y - (L^4)_+ + L^4 \] \cr
& = - \left\{ L,\, \[ \partial_y - (L^4)_+ , \, L\] \right\} =0
{}~~,  \cr
F_{y t} &= \[ \partial_y - (L^4)_+ + L^4 , \,
\partial_t - (L^6)_+ + L^6 \] \cr
& = \[ \partial_y - (L^4)_+ , \, L^6  \] - \[ \partial_t - (L^6)_+, \,
L^4\] + \[ \partial_y - (L^4)_+ , \, \partial_t - (L^6)_+ \] =0 ~~. \cr }
\eqno(2.10) $$
We have made use of the relations
$$ \eqalign{&\fracmm{\partial L^{2p} } {\partial t_{2n}}
= \[ (L^{2n})_+, L^{2p} \] ~~, \cr
& \fracmm {\partial (L^{2n})_+} {\partial t_{2m} }
- \fracmm {\partial (L^{2m})_+} {\partial t_{2n}} = \[ (L^{2m})_+,
{}~(L^{2n})_+\] ~~, \cr }
\eqno(2.11) $$
which can be easily shown to hold for arbitrary natural numbers
$~m,\,n$~ and $p$.  Similarly we can get
{}~$F_{t\theta} = F_{x t} = 0$.
It is interesting to see that $~A_a~{\scst (a ~=~
t_{2n})}$~ can be rewritten as $~A_{t_{2n}} = (L^{2n})_-$.

	In the above formulation, we have only dealt with the embedding
into the $~D=4$~ SDSYM.  In fact, based on the more general formulation
in ref.~[11], it is expected that the embedding can be generalized to any
even-dimensional (generalized) SDSYM
theories.  However, the $~D=4$~ SDSYM is to be the most well-motivated
from the viewpoint of $~N=2$~ superstring, as its effective background.

	We mention the arbitrariness related to the identification
(2.9a).  There are infinitely many alternative identifications, such as
$$ \nabla_\theta^{(k)} \equiv L^{2k+1} ~~, ~~~~ (k = 0,\, 1, \, \cdots)
{}~~.
\eqno(2.12)$$
For the embedding into $~D=3$, we choose any arbitrary three bosonic
coordinates such as $~x\equiv t_{2(2k+1)}, \, y\equiv t_{2l},\, t\equiv
t_{2m}$, containing $~x\equiv t_{2(2k+1)}$~ out of the set
$~\{t_2,\,t_4,\,t_6,\, \cdots\}$, with other coordinates truncated.  Now we can
confirm the set
$$\eqalign {& \nabla_\theta^{(k)} \equiv L^{2k+1}~~, \cr
& \nabla_x \equiv \partial_x + A_x \equiv L^{2(2k+1)} ~~, \cr
& \nabla_y \equiv \partial_y + A_y \equiv \partial_y - (L^{2l})_+ + L^{2l}
{}~~, ~~~~ (l,\, m = 1,\, 2,\, \cdots) ~~, \cr
& \nabla_t \equiv \partial_t + A_t \equiv \partial_t -(L^{2m})_+ +
L^{2m} ~~, ~~~~(l\neq m\neq 2k+1 \neq l)~~,  \cr }
\eqno(2.13) $$
satisfies our fundamental equation (2.8) again, by the help of (2.11).
This sort of arbitrariness in our superspace, or in other words, the
arbitrariness for choosing a superspace is closely related to
the existence of infinitely many conserved charges discussed later.

\bigskip\bigskip
\bigskip\bigskip

\noindent {\bf 3.~~ Relationship with Super-Sato Equation}
\vskip 0.1in

	The vanishing superfield strength suggests that all the
potential superfields must be {\it pure-gauge} in our generalized
algebra space (ring) $~{\cal E}_{\cal S}^{(1|1)}$.  We can show
this is indeed the case, and it has more
significance, when we review the super-Sato equation [11] associated with the
super-Lax equation (2.1).

	The super-Sato equation [11] is just {\it supersymmetric}
generalization of what is called Sato equation [13] for the {\it bosonic}
Lax equation.  It has
been known that the super-Lax equation (2.1) can be solved, if and {\it only
if} there exists a bosonic super-microdifferential operator
$$W = \sum _{n=-\infty} ^ 0 w_n (x,y,t,\theta) D^n \in {\cal E}_{\cal
S}^{(1|1)} (0) ~\bigcap~ ({\cal E}_{\cal S}^{(1|1)})_0  ~~,
\eqno(3.1) $$
satisfying
$$\li{& L = W D W^{-1} ~~,
&(3.2a) \cr
& \fracmm {\partial W} {\partial t_{2n}} = (L^{2n})_+ W - W D^{2n} ~~.
&(3.2b) \cr } $$
This $~W$~ is usually called super wave operator, and (3.2b) is the
super-Sato equation [11].
In the purely bosonic system, it is known that $~W$~ can be regarded as
a gauge transformation parameter [12].  We can easily see that this is
also the case with the {\it super}-Lax equations after simple algebra.  In
fact, inserting (3.2a) into our ans{\" a}tze (2.9), and using also
(3.2b), we get
$$\li{& \nabla_\theta = W D W^{-1} ~~,~~~~
\nabla_x = W \partial_x W^{-1} ~~,
&(3.3a) \cr
& \nabla_y = \partial_y - (L^4)_+ + W D^4 W^{-1}
= W \partial_y W^{-1} ~~, \cr
& \nabla_t = \partial_t - (L^6)_+ + W D^6 W^{-1}
= W \partial_t W^{-1} ~~.
&(3.3b) \cr } $$
Note that the second equalities for $~\nabla_y$~ and
$~\nabla_t$~ hold only {\it on-shell}, namely only after the use of
the super-Sato equation (3.2b).  In these forms of $~\nabla_A$, it is
clear that $~F_{A B}$~ vanishes in this system.  To put it differently,
the vanishing superfield strength (2.8) is a
sufficient condition of the super-Sato equation (3.2b), {\it via} the
identification (3.2a).

\bigskip\bigskip
\bigskip\bigskip

\noindent {\bf 4.~~ Infinite Number of Conserved Charges~~}
\vskip 0.1in

We can also understand the infinitely many conservation laws [10]
associated with the whole hierarchy contained in the super-Lax equation
in our superspace formulation.  These conservation laws can be formulated
by the use of the arbitrariness we mentioned with (2.12).
In fact, we can define the generalized
fermionic charges [10]
$$Q_{\rm F}^{(k)} \equiv \int d x d y d\theta ~\res (L^{2k+1}) \equiv
\Tr (L^{2k+1}) = \Tr (\nabla_\theta^{(k)} ) ~~, ~~~~(k=0,\,1,\,\cdots)
{}~~,
\eqno(4.1) $$
where the symbol $~{\cal \res}$~ implies the coefficient of the
$~D^{-1}\-$term in an arbitrary super-microdifferential operator $~\G$:
$$\eqalign{&\res \G \equiv g_{-1} (x,y,t,\theta)  ~~, \cr
&\G \equiv \sum_{n=-\infty} ^m g_n(x,y,t,\theta) D^n ~~,
{}~~~~(m=1,\,2,\,\cdots) ~~. \cr }
\eqno(4.2) $$
Eq.~(4.1) is interpreted as the generalization of
$$ Q_{\rm F} = \int d x d y d\theta ~ \res (\nabla_\theta) \equiv \Tr
(\nabla_\theta) ~~,
\eqno(4.3) $$
based on the arbitrariness mentioned in (2.12).
We can show for arbitrary super-microdifferential
operators $~\G$~ and ~$\L$~ that [9]
$$\int d x d y d\theta ~\res \[ \G,\, \L \]  = \Tr \[ \G,\,\L \] = 0 ~~,
\eqno(4.4) $$
because $~\res \[\G,\, \L  \] $~ is always a
total divergence with respect to $~x,~y,~\theta$.
It is now straightforward to show for $~t\equiv t_6$~ that
$$\eqalign{ \fracmm {d Q_{\rm F}^{(k)}}{d t} & = \fracmm d {d t}
\int d x d y d\theta ~ \res (L^{2k+1})  \cr
& = \int d x d y d\theta ~\res \[ (L^6)_+,\, L^{2k+1} \] = 0 ~~. \cr}
\eqno(4.5) $$
Eq.~(4.5) holds as a sufficient condition, when all the
eigenvalues of the operator $~\nabla_\theta^{(k)}$~ are {\it isospectral},
({\it i.e.,}$~t\-$independent), as is seen from the Adler-trace expression in
(4.1).  By appropriate truncation [14] to the ~$N=1$~ SKdV system [9],
we easily see that $~Q_{\rm F}^{(0)}$~ agrees with the fermionic charge
$~\int d x\, \xi (x)$, where $~\xi(x)$~ is the fermionic component field
in the lowest flow of $~N=1$~ SKdV equation [9].  The $~Q_{\rm
F}^{(k)}$~ with higher $~k$~ seem to yield new fermionic charges.

	The existence of the infinitely many conserved charges
$~Q^{(k)}_{\rm F}$~ is the reflection of the arbitrariness in
$~k$~ in the identification $~\nabla_\theta^{(k)} \equiv L^{2k+1}$~ in our
superspace.  This feature seems universal in any super-Lax equation
with a fermionic super-Lax operator $L$.  We should also notice the
close relationship of this arbitrariness with the vanishing superfield
strength.

	We can also consider the conserved {\it bosonic} charges
$$ Q_{\rm B} ^{(k)} \equiv \int d x d y d\theta~ \res(L^{2k} )
= \Tr (\nabla_{t_{2k}}) ~~, ~~~~ (k = 1,\, 2,\, \cdots) ~~.
\eqno(4.6) $$
Here we used the general formula (2.13) for $~\nabla_{t_{2k}}$, and for
appropriate coordinates $~x\equiv t_{2(2k+1)},~y \equiv t_{2l},~ t\equiv
t_{2m}$.  After the dimensional reduction into $~D=2$~ as in ref.~[14],
eq.~(4.6) is reduced to the infinite bosonic charges for $~N=1$~ SKdV
hierarchy [9].

\bigskip\bigskip
\bigskip\bigskip

\noindent {\bf 5.~~Concluding Remarks~~}
\vskip 0.1in

	In this Letter we have shown explicitly that the super-Lax
operator for the \hbox{$~D=3$}, ~\hbox{$N=1$}~ SKP of Manin and Radul [10]
is embeddable into the $~D=4,\,N=1$~ SDSYM theory, {\it via} the super-Lax
operator embedded in our superspace.  We have also seen how the super-Sato
equation is related to the super-Lax equation in a differential geometrical
manner.  The significance of infinitely many conserved charges is also
clarified in our superspace.
This is a good explicit indication that such an embedding is universally
applicable to other super-Lax equation systems, and
thus provides a strong support that any supersymmetric integrable
model can be embedded into the $~D=4$~ SDSYM theory, which is the
supersymmetric version of the original Atiyah's conjecture [1] for purely
bosonic integrable models.

	Understanding of the relationship between the
super-Lax equation and the corresponding super-Sato equation in our
embedding principle has elucidated the geometrical significance of
the latter in terms of vanishing superfield strength in superspace.
It is not a coincidence that there exists a super wave operator $~W$~ for
each super-Lax operator $~L$, because the former is nothing but a
supersymmetric gauge transformation for a {\it pure-gauge} potential
superfield.  Even though any system that has only pure gauge potential field is
{\it not} usually interesting physically except for non-trivial topology,
we have seen that this observation is {\it not} the case with the
integrable systems in lower-dimensions, which provide many physically
interesting models.

	The new ingredient in this paper we emphasize is the {\it
superfield} strength formulation related to the $~N=2$~ superstring
theory applied to the super-Lax operator, which has
not been presented in the past literature to our knowledge.\footnotew{A
prototype geometrical interpretation can be found in refs.~[10,11], but
no close relationship with the vanishing superfield strength in superspace,
or with the more fundamental $~N=2$~ superstring theory was pointed out.}
It has been known that the {\it bosonic} Lax equation can be reformulated in a
zero-field strength system, either for the finite-dimensional graded
Lie groups [14] or in the infinite-dimensional generalization [12].
Our result that the {\it super}-Lax equation is closely related to the SDSYM
theory in $~D=4$~ {\it via} super-Sato equation for pure-gauge
superpotential signals the fundamental importance of the SDSYM theory
[3,4] for supersymmetric integrable models.

	We have seen that the infinitely many conserved
charges associated with the super-Lax equation is the reflection of the
arbitrary fermionic operators $~\nabla_\theta^{(k)}\equiv L^{2k+1}$~
in our superspace.  It also corresponds to the alternative choices
of three bosonic coordinates out of infinitely many time coordinates [11] in
the whole SKP hierarchy.

	The features of the
super-Lax equation we utilized in our embedding are pretty common to
any super-Lax hierarchy with a fermionic Lax operator.  In particular, the
patterns we have seen in sections 2 through 4 are
universal, not limited to the SKP hierarchy.  This implies that any
super-Lax operator in the super-Lax equation (2.1) can follow the same
embedding procedure into the $~D=4$~ SDSYM theory, {\it via} vanishing
superfield strength (2.8) in superspace.

	We have so far stressed the relationship between the super-Lax
equation and the \hbox{$~D=4$}
SDSYM.  As some readers may have
already noticed, our basic equation (2.8) in superspace has more to do
with supersymmetric Chern-Simons theory of vanishing superfield
strength in $~D=3$.  It is no wonder that the three systems, namely the
super-Lax equation [10,11], the SDSYM theory in $~D=4$~ [3,4], and
supersymmetric Chern-Simons theory in $~D=3$~ [15] have such close mutual
relationships, which are dictated in terms of differential geometry in
mathematics, and are motivated by $~N=2$~ superstring theory [2-4] as their
underlying ``master theory''.

\bigskip\bigskip

The author is grateful to S.J\.Gates, Jr.~for valuable suggestions.

\bigskip\bigskip

\vfill\eject

\refs
\small

\Item{[1]} M.F.~Atiyah, unpublished;
\Item{  } R.S\.Ward, Phil.~Trans.~Roy.~Lond.~{\bf A315} (1985) 451;
in ``{\it Field Theory, Quantum Gravity and Strings}'', ed.~H.J.~de Vega
and N.~Sanchez, Springer Lecture Notes in Physics, Vol.~{\bf 246} (1986).

\Item{[2]} H.~Ooguri and C.~Vafa, \mpl{5}{90}{1389};
\np{361}{91}{469}; \ibid{367}{91}{83};
\Item{  } H.~Nishino and S.J.~Gates, Jr., \mpl{7}{92}{2543};
\Item{  } N.~Berkovitz and C.~Vafa, Harvard - King's College preprint,
HUTP-93/A031, KCL-TH-93-13 (Oct.~1993).

\Item{[3]} W.~Siegel, \pr{47}{93}{2504}.

\Item{[4]} S.V.~Ketov, S.J.~Gates, Jr.~and H.~Nishino, \pl{308}{93}{323};
\Item{  } H.~Nishino, S.J.~Gates, Jr. and S.V.~Ketov,
\pl{307}{93}{331};
\Item{  } S.J.~Gates, Jr., H.~Nishino and S.V.~Ketov,
\pl{297}{92}{99};
\Item{  } S.J.~Ketov, H.~Nishino and S.J.~Gates, Jr., \np{393}{93}{149}.

\Item{[5]} S.J.~Gates, Jr.~and H.~Nishino, \pl{299}{93}{255}.

\Item{[6]} H.~Nishino, Maryland preprint, UMDEPP 93--145 (Feb.~1993), to
appear in Phys.~Lett.~B.

\Item{[7]} H.~Nishino, \pl{309}{93}{68}.

\Item{[8]} H.~Nishino, Maryland preprint, UMDEPP 93--213 (June 1993), to
appear in Phys.~Lett.~B.

\Item{[9]} P.~Mathieu, \pl{203}{88}{287}; \jmp{29}{88}{2499}.

\Item{[10]} Y.~Manin and A.O.~Radul, \cmp{98}{85}{65}.

\Item{[11]} K.~Ueno and H.~Yamada, Lett.~Math.~Phys.~{\bf 13} (1987) 59;
\Item{   } Y.~Ohta, J.~Satsuma, D.~Takahashi, and T.~Tokihiro,
Prog.~Theor.~Phys.~Suppl.~{\bf 94} (1988) 210.

\Item{[12]} J.~Schiff, Princeton preprint, IASSNS--HEP--92/34 (Oct.~1992).

\Item{[13]} M.~Sato and Y.~Sato, in ``{\it Non-Linear Partial Differential
Equations in Applied Science}'',
\Item{  } ed.~H.~Fujita, P.D.~Lax and G.~Strang,
(Kinokuniya/North Holland, Tokyo, 1983) p.~259.

\Item{[14]}  J.~Barcelos-Neto, A.~Das, S.~Panda, and S.~Roy,
\pl{282}{92}{365}.

\Item{[15]} H.~Nishino and S.J.~Gates, Jr., \ijmp{8}{93}{3371}.

\end{document}